**New cloud morphologies discovered on the Venus's night during Akatsuki**

J. Peralta[1], A. Sánchez-Lavega[2], T. Horinouchi[3], K. McGouldrick[4], I. Garate-Lopez[2], E. F. Young[5], M. A. Bullock[5], Y. J. Lee[6], T. Imamura[6], T. Satoh[1] and S. S. Limaye[7].

(1) Institute of Space and Astronautical Science (ISAS/JAXA), Kanagawa, Japan.

(2) Grupo de Ciencias Planetarias, Departamento Física Aplicada I, E.T.S. Ingeniería, UPV/EHU, Spain.

(3) Faculty of Environmental Earth Science, Hokkaido University, Sapporo, Japan.

(4) Laboratory for Atmospheric and Space Physics, Univ. of Colorado Boulder, Boulder, USA.

(5) Southwest Research Institute, Boulder, CO 80302, USA.

(6) Graduate School of Frontier Sciences, University of Tokyo, Japan.

(7) Space Science and Engineering Center, University of Wisconsin-Madison, Madison, USA.

**Abstract:** During the years 2016 to 2018, the instruments Akatsuki/IR2 (JAXA) and IRTF/SpeX (NASA) acquired a large set of images at 1.74, 2.26 and 2.32 µm to study the nightside mid-to-lower clouds (48−60 km) of Venus. Here we summarize the rich variety of cloud morphologies apparent in these images: from frequent wave packets and billows caused by shear instabilities, to features reported decades ago like the circum-equatorial belts, bright blotches and equatorial troughs, and previously unseen features like dark spots, sharp dark streaks at mid latitudes and fully-developed vortices.

**1. Introduction.**

Venus is permanently shrouded by a thick layer of clouds mainly composed of sulphuric acid droplets and covering heights within 48–74 km (Knollenberg & Hunten, 1980; Ignatiev et al. 2009). The upper clouds (~60−74 km) can be observed on both day and night sides with ultraviolet–violet and infrared wavelengths, respectively (Peralta et al. 2017a). The dayside clouds have been studied for decades, providing valuable constrains to chemical and convective processes and the atmospheric circulation (Marcq et al. 2018; Lefèvre et al. 2016; Sánchez-Lavega et al. 2017). They exhibit waves and varied morphologies strongly influenced by advective and shearing processes (Schinder et al. 1990; Titov et al. 2012; Patsaeva et al. 2015): waves like the Y-feature, circum-equatorial belts (large equatorial wave packets with wave fronts zonally oriented), bow-like and mesoscale waves (Belton et al. 1976; Rossow et al. 1980; Peralta et al. 2008), convective cells (Rossow et al. 1980), spiral streaks, bright bands and polar vortices (Titov et al. 2012; Muto & Imamura, 2017; Limaye et al. 2018). In infrared, the upper clouds show the polar vortices with detail (Garate-Lopez et al. 2013), abundant stationary features linked to surface elevations (Peralta et al. 2017b; Kouyama et al. 2017), fast filaments and shapes reminiscent of shear instabilities (Peralta et al. 2017b).

The deeper clouds of Venus (48−60 km) can be studied at the infrared windows at 1.74, 2.26 and 2.32 µm (Peralta et al. 2017a), which allow sensing the lower clouds' variable opacity to the deeper thermal emission, with thicker clouds observed as dark areas and thin clouds as bright ones (Allen & Crawford, 1984; Hueso et al. 2012; McGouldrick et al. 2012). Ground-based observations revealed that the global opacity exhibits occasional hemispherical asymmetry, with mid-latitudes (40°−60°) usually occupied by zonal bright bands (i.e. with low opacity) while between 40°N−40°S a persistent planetary-scale homogeneous dark cloud (high opacity) alternates with a more heterogeneous area of mixed dark and bright markings of smaller spatial scales of 400−1,000 kilometres (Crisp et al. 1991). During 2006−2008, images from the instrument VIRTIS onboard Venus Express (VEx) permitted to visualize with unprecedented spatial resolution the lower clouds at high latitudes, revealing the deeper structure of the southern polar vortex (Garate-Lopez et al. 2013), recurrent wave packets (Peralta et al. 2008) or opacity holes (Hueso et al. 2012). After its orbit insertion in 2015 December (Nakamura et al. 2016), JAXA's Akatsuki orbiter started to observe with great detail the deep clouds of Venus at low latitudes using the 2-µm camera (IR2) and filters 1.74, 2.26 and 2.32 µm (Satoh et al. 2017), and reported new cloud morphologies like sheared features or sharp opacity discontinuities at the equator (Satoh et al. 2017; Horinouchi et al. 2017; Limaye et al. 2018; Kashimura et al. 2019). Here we present a first morphological classification of the lower clouds and report the discovery of new phenomena like low-latitude vortices, sharp dark streaks encircling the globe or dark spots, and patterns that had been only observed at the upper clouds.

## 2. Methods

A total of 1,671 IR2 images between March and November of 2016 were examined to study the night lower clouds' morphology of Venus. The equatorial orbit of Akatsuki has a period of ~10 days and apoapsis/periapsis of ~360,000 km and 1,000−8,000 km, and the spatial resolution of IR2 images comprised from 74−12 km per pixel (off-pericenter observations) to 1.6−0.2 km (pericentric) (Nakamura et al. 2016). We used images taken at 2.26 µm, since these were less affected by the light pollution spreading from the saturated dayside of Venus (Satoh et al. 2017). Since IR2 ceased observations in December 2016 (Peralta et al. 2018), this study was complemented with 2.3-µm ($K_{cont}$) images with the instrument SpeX at NASA's IRTF telescope in Hawaii (Rayner et al. 2003) taken during January-February of 2017 (>700 images, spatial resolution 65−40 km/pixel) and November-December of 2018 (>3,600 images, 29−50 km/pixel).

The IR2 and SpeX images were processed (brightness/contrast adjustment, unsharp-mask sharpening and adaptive histogram equalization to enhance local contrasts) and projected onto equirectangular coordinates with the IDL software described by Peralta et al. (2018). The same software was used to measured clouds' speeds and other of their properties (see Table 1): length/width, orientation relative to lines of constant latitude (parallels), horizontal wavelength, position (latitude/longitude and local hour) and speeds relative to the zonal wind. Features were classified into 9 categories: *vortices* and *Kelvin-Helmholtz (KH) billows* (Fig. 1A-E), *shear waves* (1D-I), *dark spots* (1E,1J,1K), *dark streaks* (2A-D), *wave packets* and *wavy boundaries* (2E-G), *circum-equatorial belts* or *"CEBs"* (2H-J), *bright troughs* (2L-O) and *bright blotches* (2K). Some of these categories ("CEBs", "wave packets" …) can be also found on Venus's upper clouds.

**3. Clouds reminiscent of shear waves, KH billows, vortices and dark spots.**

The boundary between mid-latitude bright zonal bands and the equatorial dark band on the nightside frequently display isolated inclined stripes ("billows") and/or vortices (Drazin, P. G., 2003) with sizes ranging 90−4,440 km (see Figs. 1A−1B and 1D−1E), although "mushroom" shapes (Fig. 1C) and more complex ones (Fig. 1F, inside oval marking) are also possible. This contrasts with the elongated tilted streaks and bands with smooth contours of the dayside upper clouds (Titov et al. 2012; Limaye et al. 2018; Peralta et al. 2017b). Other times, trains of billows and/or vortices reminiscent of Kelvin-Helmholtz instabilities or gravity-shear waves (Lyulyukin et al. 2015) (Fig. 1, panels F−I) are observed with wavelengths ranging 65−2,575 km, lengths of 690–6,590 km and variable orientation relative to the parallels of latitude (-43º to +31º). Less recurrently, smaller trains of billows appear at high latitudes and even close to the southern polar vortex, as seen in VEx/VIRTIS images (Figs. 1H and 1I). Figures 1D and 1E show the first evidence of fully-developed vortices on the clouds of Venus (marked with arrows), exhibiting sizes ranging from 420 to 760 km. None of these features exhibit a clear dependence with the geographical longitude or the local time.

*KH billows* and *vortices* can keep a coherent shape for more than 36 hours (see Table 1) and no detectable vorticity even with images separated by more than 6-8 hours. Low-latitude billows exhibit shapes compatible with clockwise spin on the northern hemisphere (Fig. 1B) and counter-clockwise on the southern one (Fig. 1A), with March 22 and July 11 presenting exceptions to this rule. This frame of reversed spins on both sides of the equator is consistent with dynamical instabilities of barotropic nature (Fritts & Rastogi, 1985; Markowski & Richardson, 2010). We show that the Rayleigh's criterion for barotropic instabilities (Vallis, 2006) is met for the intermittent equatorial jet of Venus. In Venus's cyclostrophic regime, a frequency derived from the centrifugal terms of the momentum equations plays the role of the Coriolis effect and fulfils the $\beta$-plane approximation with $\beta \sim -1.8 \cdot 10^{-12} \, s^{-1} m^{-1}$ for the lower clouds (Peralta et al. 2014a; 2014b). The jet measured by Horinouchi et al. (2017, Fig. 2b) displays four latitude regions (35ºN, 5ºN, 20ºS and 25ºS) where the meridional shear of the zonal speed changes its sign. According to our calculations, ($\beta - \partial^2 U/\partial y^2$) adopts the values $1.1 \cdot 10^{-11}$, $-2.2 \cdot 10^{-11}$, $3.4 \cdot 10^{-11}$ and $-3.3 \cdot 10^{-11} \, s^{-1} m^{-1}$ around these latitudes, therefore changing its sign within the domain where billows and vortices appear. Conditions for barotropic instabilities were also proven with radio-occultation (Piccialli et al. 2012), while the vertical shear of the zonal wind ($|\partial U/\partial z| \sim 10^{-3} \, s^{-1}$ against $|\partial U/\partial y| \sim 10^{-5} \, s^{-1}$; Peralta et al. 2017b) or convective instabilities (Tellmann et al. 2009) cannot be ruled out as sources.

In opposition to presence of bright spots or "holes" linked to regions with smaller optical thickness (Hueso et al. 2012; Horinouchi et al. 2017), the IR2 images also display numerous and recurrent dark spots with size ranging 270–1,430 km, radiance dropping ~30% inside, and sometimes surrounded by a "ring" of bright clouds (Fig. 1, panels A, F, J and K). The dark spots display a major occurrence within geographical longitudes 10º–90º and shorter life, even vanishing in less than 4 hours (see Table 1).

**4. Sharp dark streaks, wave packets and wavy boundaries.**

The spatial resolution of the IR2 images also permits to resolve sharper cloud features like the *sharp dark streaks*, reminiscent of the spiral streaks reported on the ultraviolet dayside albedo of the cloud tops at southern polar latitudes (Belton et al. 1976; Rossow et al. 1980; Titov et al. 2012). They consist on narrow bands (widths from 40 to about 300 km) usually extending several thousands of kilometres (>9,800 km in some cases) along the bright mid-latitude bands (see Figs. 2A−2D) and causing a decrease of the radiance of up to ~40% (Fig. 2C). They are slightly tilted relative to the geographical parallels (see Table 1) and they encircle the planet towards the East, gradually approaching lower latitudes until vanishing between 20°−40°. Except during the month of October, the *dark streaks* were a recurrent feature on Venus during the year 2016, preferentially showing up on the northern hemisphere, although they can be also present simultaneously on both hemispheres (Fig. 2B) and, in rare cases, at equatorial latitudes as recently observed with IRTF observations (Fig. 2D). Preliminary cloud tracking measurements with IR2 images do not allow to confirm whether these sharp dark streaks are caused by mid-latitude jets as suggested for the dayside spirals by Titov et al. (2012). The dark streaks could be also the phase distortion of an equatorially-trapped wave with large meridional e-folding decay, as shown by Peralta et al. (2015).

Pericentric images from IR2 have better spatial resolution and they reveal an equatorial region rich in *wave packets* with properties like those identified during the VEx mission and classified as gravity waves (Peralta et al. 2008; 2014a): wavelengths ranging from 70 km (see Fig. 2F) to 340 km (Fig. 2E), packet lengths ranging 360–1,380 km and a variable orientation (see Table 1). Figure 2G also displays that the boundary between the dark equatorial and bright mid-latitude bands (0°−30°S), and a small area at higher latitudes (30°S−45°S) are subject to wavy disturbances of larger spatial scale, with wavelengths of 5,970 km (northern hemisphere) and 4,370 km (southern hemisphere). In accordance with analyses of VEx data (Peralta et al. 2017b), no stationary waves were found either on the lower clouds after analysing the full dataset of the IR2 images.

**5. Circum-equatorial belts (CEBs), the bright blotch and equatorial trough.**

Observations along September–October 2016 confirm a long-term recurrence (decades) for certain phenomena. An example are the *circum-equatorial belts* or *CEBs*, firstly observed on the dayside upper clouds from ultraviolet images during NASA's Mariner 10 flyby (Belton et al. 1976). The CEBs consist on large wave packets with wave fronts (i.e. the lines of constant phase of the wave) oriented zonally (see Figs. 2H−2J). During the flyby of Mariner 10, the *CEBs* of the dayside upper clouds exhibited horizontal wavelengths of ~500 km, packet lengths and widths about 5,000 and 50−150 km, southward phase speeds of ~20 m s$^{-1}$ and a life span ranging 0.5−1.5 days. The *CEBs* on the night lower clouds during the Akatsuki mission display horizontal wavelengths ranging 90−400 km, packet lengths of 1,200−6,620 km and variable width of 50−2,920 km, no apparent meridional drift and a life span that can exceed 1 day (see Table 1).

Another example is the formation of a large *bright blotch* or spot of 4,460 km at about 20°S during October 2016 (see Figure 2K). Crisp et al. (1991) reported a similar though smaller (~2,000 km) bright spot at 16ºN in their ground-based observations in January 29 and, again, in February 9 of 1990. The IR2 observations during October 2016 also confirm that the bright bands at mid-latitudes can invade equatorial latitudes until merging and forming a *bright trough* (Fig. 2, panels L−O) and a clear *CEB* on its Eastern side (Figs. 2N and 2I). This *trough* extended from 25°S to 35°N, presents different orientations in each hemisphere (Fig. 2N) and can keep coherent up to ~10 days. Despite their worse spatial resolution, the observations by Crisp et al. (1991) also display the formation of this bright equatorial *trough*, suggesting its recurrent nature.

## **6. Conclusions.**

We performed a detailed characterization and classification of the morphology of the nightside lower clouds of Venus using images from Akatsuki/IR2 and IRTF/SpeX. While some cloud patterns are new, others had been observed only on the dayside, thus providing hints about how the same phenomenon affects both clouds' opacity and albedo. Low-latitude clouds are rich in mesoscale *wave packets* and *black spots*, while the frequent *KH billows* and episodical *vortices* could be key to explain the generation of waves and constrain the momentum flux due to eddies, and they are probably caused by the combined effect of the *meridional shear* associated to the intermittent equatorial jet (Horinouchi et al. 2017) and the *centrifugal factor* that plays the role of the Coriolis effect in the cyclostrophic regime of Venus (Peralta et al. 2014b). Finally, the presence of *sharp dark streaks* encircling the planet and the convergence of the mid-latitude bands merging into a prominent *bright trough* at the equator imply that the lower clouds are more complex than expected and imply new challenges for the numerical models.

**ACKNOWLEDGEMENTS:**

JP acknowledges JAXA's International Top Young Fellowship. ASL thanks the Spanish MINECO project AYA2015-65041-P with FEDER, UE support, Grupos GV IT-765-13 and UPV/EHU program UFI11/55. IGL was supported by the Basque Government's Doctoral Research Staff Improvement Programme. KM and SSL thanks NASA Grants NNX16AC80G and NNX16AC79G. All authors acknowledge the Akatsuki team.

| Cloud Feature | Nº of Cases | Length (km) / Width (km) | Wave Length (km) | Orientation (relative to parallels) | Latitudes / Longitudes / Local Times | U−U₀ (m/s) | V (m/s) | Life Span (hours) | Cause? | Notes |
|---|---|---|---|---|---|---|---|---|---|---|
| Billows & Vortices (Fig. 1A-E) | 35 | 89±18 to 4,443±53 | — | — | 37ºN–73ºS / 0º–360º | 0±3 | -3±7 | Min: < 21  Max: > 36 | Dynamical instability? Waves? | VEx observations included. No dependence with Longitude or LocalTime |
| Shear waves (Fig. 1D-I) | 23 | 687±25 to 6,585±18 | 65±11 to 2,575±11 | -43º to +31º | 19h–06h | 2±5 | 10±7 | Min: < 21  Max: > 21 | | |
| Dark Spots (Fig. 1E,1J,1K) | 46 | 270±43 to 1,432±69 | — | — | 43ºN–51ºS / 0º–360º / 20h–04h | 0±7 | 4±8 | Min: < 4  Max: > 24 | ? | Major occurrence at Longitudes 10º–90º |
| Sharp Dark Streaks (Fig. 2A-D) | 10 | >3,469±26 to >9,800±48 / 36±14 to 303±48 | — | -38º to +11º | 61ºN–35ºS / 6º–360º / 18h–06h | 0±3 | -6±4 | > 48 | Wave? | Sometimes concentric streaks separated by ~600 km. Fig. 2B |
| Wave Packets (Fig.2E-F) | 13 | 356±18 to 1,381±37 / 166±7 to 3,327±37 | 66±8 to 335±37 [Avg: 149±75] | -87º – +89º [Avg: +12±43] | 26ºN–43ºS / 67º–265º / 19h–05h | -4±9 | -3±6 | Min: < 23  Max: > 30 | Waves | — |
| Wavy Boundaries (Fig. 2G) | 2 | >9,407±43 / 945±43 to 1,163±43 | 5,967±43 & 4,367±43 | -21º & +12º | 52ºN–22ºN & 7ºS–28ºS / 44º–135º / 18h–01h | -1±7 | -3±7 | > 30 | Wave? | Meridional oscillations 940−1200 km |
| Circum-equatorial Belts (CEB) (Fig. 2H-J) | 6 | 1,203±15 to 6,619±21 / 48±21 to 2,919±80 | 95±21 to 404±80 | -88º – +90º | 47ºN–28ºS / 74º–160º & 243º–303º / 18h–05h | -3±4 | 0±8 | Min: < 10  Max: > 26 | Wave? | March 25, September 5, October 7, 11-12, 26 |

| Bright Trough (Fig. 2L-O) | 4 | 3,578±62 to 8,605±81<br><br>262±73 to 1,296±62 | — | North: +42± 1°<br>South: -45±16° | 35°N–25°S<br>155°–214°<br>21h–01h | -3±6 | 0±8 | Min: < 23<br>Max: > 4 | ? | Only during October 2016. Anomalous orientation in Fig. 2M |
| --- | --- | --- | --- | --- | --- | --- | --- | --- | --- | --- |
| Bright Blotch (Fig. 2K) | 1 | 4,456±80 | — | — | 5°S–37°S<br>137°–181°<br>21h–23h | -1±11 | 1±11 | > 2 | ? | October 3 only |

**Table 1:** Summary of main Venus nightside cloud features found in the Akatsuki/IR2 images, and average values and/or range of their main properties. In those cases of stretched shapes, the length and width are presented, while in the rest of the cases the size is defined with a single value. When the cloud features exhibit a wavy pattern, the corresponding horizontal wavelength is provided. When the shape of a type of cloud feature presents a strong degree of asymmetry, the orientation is given as the angle of its longest dimension relative to the geographical parallels (negative when clockwise). In the case of wave packets, the orientation is given for the approximated wave vector perpendicular to the wave fronts. The intrinsic zonal speed relative to the background zonal wind (U−$U_0$) and the meridional speed (V) are also provided.

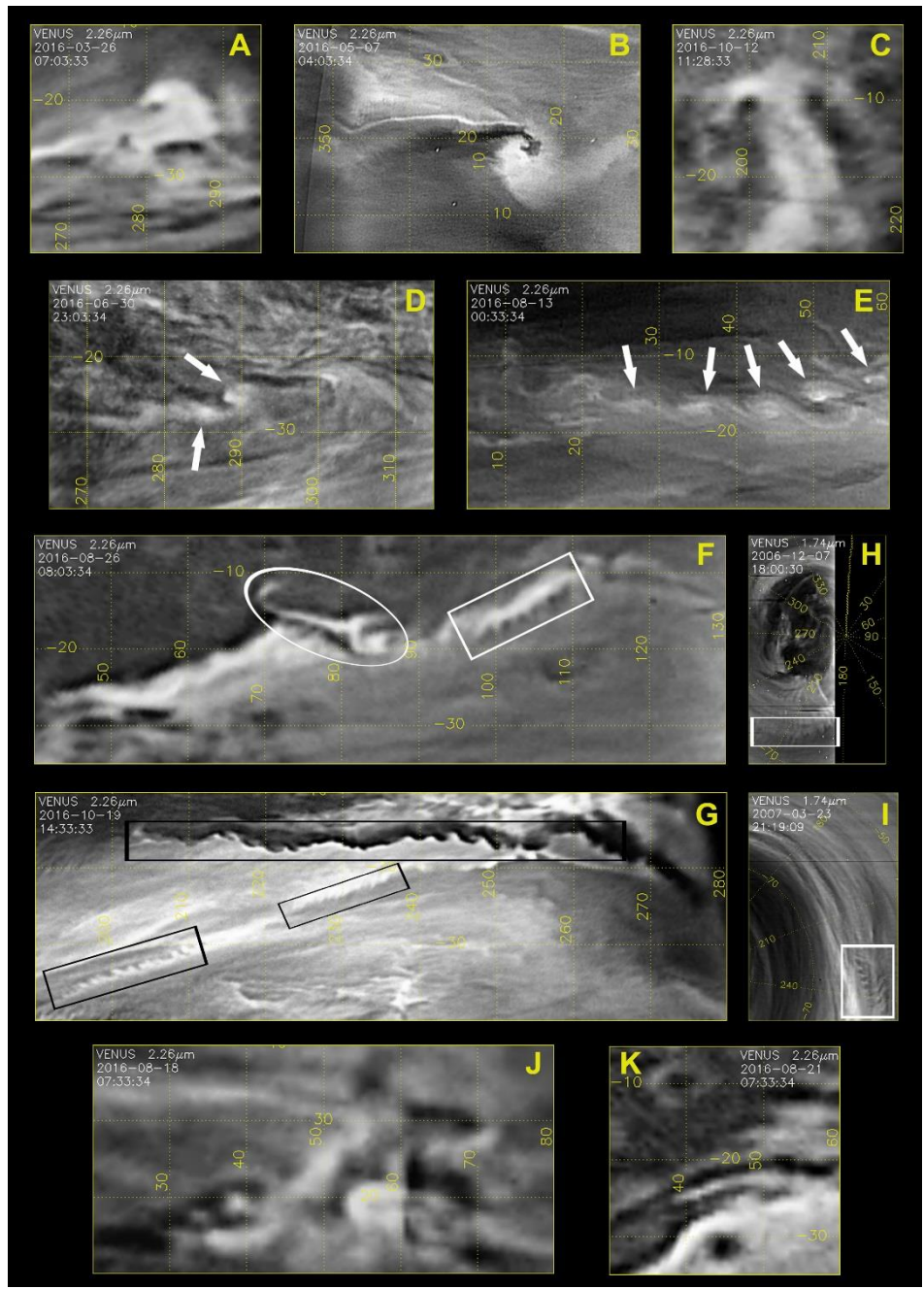

**Figure 1:** Examples of isolated billows (A−C and I), shear waves (F−I, marked with rectangles), vortices (D and E, marked with arrows) and black dots (F, J and K) on the nightside lower clouds of Venus from Akatsuki IR2/2.26-µm images (A−G and J−K) and VEx VIRTIS-M/1.74-µm images (H−I). All the images were processed as explained in the article. Except for the case of VIRTIS-M, the rest of the image were geometrically projected onto equirectangular coordinates with a resolution of 0.1° (~11 km) per pixel. The dates and times are also displayed in each image. The original spatial resolution (per pixel) of the images in the areas of interest are: (A) 49 km, (B) 14 km, (C) 80 km, (D) 32 km, (E) 34 km, (F) 50 km, (G) 19 km, (H) 22 km, (I) 18 km, (J) 89 km, and (K) 78 km.

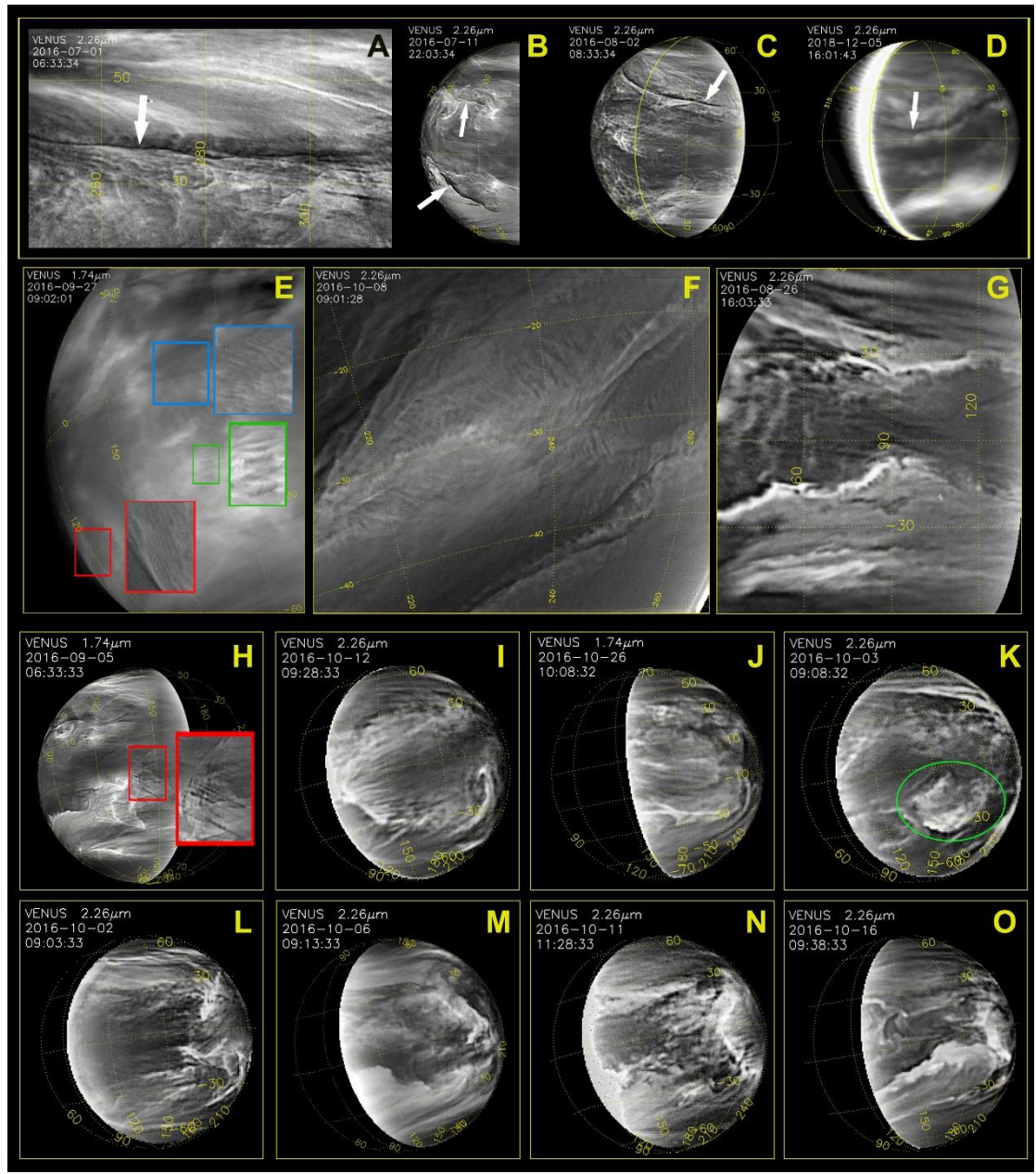

**Figure 2:** IR2/2.26-µm images displaying on the nightside lower clouds of Venus the dark streaks (A−D), wave packets (E−F), wavy boundaries (G), circum-equatorial belts or "CEBs" (H−J), the bright blotch (K), and the bright equatorial wall (L−O). The date and time are displayed in each image, and all the cases display the original perspective from the Akatsuki orbiter except for the equirectangular projections shown in (A) and (G) and the azimuthal equidistant projection in (D) (northern/southern hemispheres are above/below, respectively). The approximate spatial resolution (per pixel) in the areas of interest are: (A) 13 km, (B) 20 km, (C) 29 km, (D) 47 km, (E) 13 km, (F) 7 km, (G) 55 km, (H) 19 km, (I) 81 km, (J) 82 km, (K) 81 km, (L) 80 km, (M) 64 km, (N) 75 km, and (O) 74 km.